# Using chronoamperometry to rapidly measure and quantitatively analyse rate-performance in battery electrodes

Ruiyuan Tian,[1,2] Paul J. King,[3] Joao Coelho,[2,4] Sang-Hoon Park,[2,4] Dominik V Horvath,[1,2] Valeria Nicolosi,[2,4] Colm O'Dwyer,[2,5] Jonathan N Coleman[1,2]*

[1]*School of Physics, Trinity College Dublin, Dublin 2, Ireland*

[2]*AMBER Research Center, Trinity College Dublin, Dublin 2, Ireland*

[3]*Efficient Energy Transfer Department, Bell Labs Research, Nokia, Blanchardstown Business & Technology Park, Snugborough Road, Fingal, Dublin 15, Ireland*

[4]*School of Chemistry, Trinity College Dublin, Dublin 2, Ireland*

[5] *School of Chemistry, University College Cork, Tyndall National Institute, and Environmental Research Institute, Cork T12 YN60, Ireland*

*colemaj@tcd.ie (Jonathan N. Coleman); Tel: +353 (0) 1 8963859.

ABSTRACT: For battery electrodes, measured capacity decays as charge/discharge current is increased. Such rate-performance is important from a practical perspective and is usually characterised via galvanostatic charge-discharge measurements. However, such measurements are very slow, limiting the number of rate experiments which are practical in a given project. This is a particular problem during mechanistic studies where many rate measurements are needed. Here, building on work by Heubner at al., we demonstrate chronoamperometry (CA) as a relatively fast method for measuring capacity-rate curves with hundreds of data points down to C-rates below 0.01C. While Heubner et al. reported equations to convert current transients to capacity vs. C-rate curves, we modify these equations to give capacity as a function of charge/discharge rate, R. We show that such expressions can be combined with a basic model to obtain simple equations which can fit data for both capacity vs. C-rate and capacity vs. R. We demonstrate that these equations can accurately fit experimental data within normal experimental ranges of rate. However, we also observe that, at high rates, the curves obtained from CA deviate from the normal behaviour showing a new, previously unobserved, decay feature. We associate this feature with the very early part of the current transient where electronic motion dominates the current. Using a simple model, we show that the dependence of the high-rate time constant on electrode thickness can be linked to electrode conductivity.

## INTRODUCTION

Lithium ion batteries are becoming increasingly important for a range of applications including electric vehicles, grid scale energy storage and portable electronic devices.[1-2] Alongside factors such as energy density and stability, rate-performance is an important metric for battery operation as it determines factors such as power deliver and charging time. Specifically, advancing rate-performance has become increasingly important as the development of electric vehicles becomes more urgent.

However, although basic rate-performance experiments are ubiquitous in papers on electrode materials,[3-5] it is relatively rare to see quantitative analysis of rate-performance data. For example, we have recently shown that a considerable amount of information can be obtained by analysing the dependence of rate-performance on electrode thickness.[6-7] Alternatively, one can imagine characterising rate-performance as one varies particle size, electrode conductivity, film density, or porosity.[7-9] Although such measurements are sometimes made,[10-13] they are much rarer than their usefulness would imply. The reason for this is simple: rate-performance experiments are very slow, particularly if data is required at low rate. For example, measuring five galvanostatic charge-discharge cycles at 0.1 C takes approximately 100 hours. Although, measurements at higher C-rates are commensurately faster, collecting an extensive data set is extremely slow.

This is a significant problem for a number of reasons. Firstly, the long times associated with rate measurements prohibit studies which require measurements on multiple samples. This limits the type of experiments described above, rendering very extensive studies impractical in most research labs. However, even if only one sample is being studied (say a new electrode material), the slowness of rate measurements typically limits the number of different rates that are tested, limiting data density. This limits the accuracy of quantitative analysis, for example the fitting of capacity-rate data.[7] Finally, the maximum performance of an electrode material can only be fully assessed in the limit of low rate as it is the capacity at low rate which should be compared to the theoretical capacity.[14] However, because low-current measurements can be extremely time-consuming, there are generally practical limits to low-rate testing. For these reasons, what is needed is a faster way to measure capacity versus rate data which yields a high density of data points down to low rates.

Very recently, Heubner et al. proposed such a solution.[15] They showed that chronoamperometry (CA) can be used to produce capacity-rate data which matches closely to that produced by traditional galvanostatic charge-discharge (GCD) experiments. Not only is CA considerably faster that GCD but it yields a much richer data set, with capacity data for typically hundreds of rate values down to extremely low rates. We believe this paper represents an important breakthrough, although it does not seem to have received the attention it deserves.

Here we build on the work of Heubner et al., further developing CA as a method for making rate measurements. In addition to further characterising this method, we develop simple equations which allows the fitting of capacity vs. rate data, whether expressed in terms of C-rate (i.e. current) or charge/discharge rate, R. This highlights the differences between analysis in terms of C-rate or R and suggests advantages to using the latter. Finally, we show that capacity vs. rate measurements via CA are considerably richer than those by traditional methods and contain hitherto unreported structure at high rate.

## RESULTS AND DISCUSSION

### *Extracting capacity-rate data from current transients*

Experimentally, CA involves applying a potential step to the electrode (i.e. switching the potential rapidly from an initial value to a final value which is then held constant) and then measuring the resultant current transient. The insight of Heubner et al. was to propose equations allowing the conversion of the current transient data (i.e. I(t)) to capacity-rate data.

In essence, Heubner et al. proposed that the I(t) data could be converted to C-rate via[15]

$$\text{C-rate} = R_C = \frac{I/M}{\int_0^{\infty}(I/M)dt} \tag{1}$$

where we normalise current to electrode mass, M. Here, the upper limit of integration of t=∞ in the denominator is important. This means the specific current is normalised to the final specific charge (i.e. that after a very long charging time). Because the current transient decays with time, this final charge is achieved after the current has become very small, i.e. when the reaction rate has become very slow. Thus the denominator in equation (1) is equal to the experimental specific capacity in the limit of low rate. We note that this definition differs slightly from the traditional definition of C-rate:

$$\text{C-rate} = R_C = \frac{I/M}{(Q/M)_{Th}} \tag{2}$$

where $(Q/M)_{Th}$ is the theoretical specific capacity. Essentially equation (1) is normalised to the actual capacity at very low rate while equation (2) is normalised to the theoretical capacity. In the ideal case, where electrodes display near-theoretical capacities at low-rate, these definitions will be virtually identical.

In addition, Heubner et al. defined the capacity relative to its maximum, low-rate value as

$$Q/Q_{low-rate} = \frac{\int_0^t (I/M)dt}{\int_0^\infty (I/M)dt} \qquad (3)$$

Using these equations, both $Q/Q_{low-rate}$ and C-rate can be extracted from the current transient yielding capacity-rate plots in a rapid manner. Experimentally, we estimate that this method is ~×3 faster than traditional GCD methods. We feel that the work of Heubner et al represents a significant breakthrough. However, we also believe that their approach can be modified in a subtle way that has a significant impact.

We have recently proposed that, rather than using C-rate, there are advantages in calculating the rate slightly differently. We have proposed calculating the charge/discharge rate, R, by normalising the current, not to the theoretical specific capacity, but to the actual measured specific capacity at a given current:[7]

$$R = \frac{I/M}{(Q/M)_E} \qquad (4)$$

where the subscript "E" represents the measured experimental capacity rather than the theoretical value. In this way, R is related to the actual charge/discharge time, rather than some idealised time. As a result, we have shown previously that data plotted as capacity vs. R can be analysed quantitatively via simple mechanistic models.[7]

In CA, the expression for R is slightly different to (1). The difference between experimentally accessible capacity at a given rate and the theoretical capacity which might be achieved at very low rate, can be expressed by integrating the current only to a specific time, t:

$$R = \frac{I/M}{\int_0^t (I/M)dt} \qquad (5)$$

In addition, as with Heubner et al., we define the capacity via the integrated current (2). However, we see no reason to normalise it to its low rate value and propose that the absolute specific capacity, Q/M, can be found from:

$$Q / M = \int_0^t (I / M) dt \tag{6}$$

We can then use equations (5) and (6) to generate capacity-rate curves from the current transient data.

*Measured CA data*

To test this, we prepared composite electrodes based on five common lithium storing cathode and anode materials: $LiNi_{0.33}Mn_{0.33}Co_{0.33}O_2$ (NMC111), graphite, $LiNi_{0.815}Co_{0.15}Al_{0.035}O_2$ (NCA), micron-Si/Graphite (μSiGr) and $Li_4Ti_5O_{12}$ (LTO). In each case, the mass loading was chosen to give an areal capacity close to 4 mAh/cm$^2$ (see methods for details) For each electrode we measured (see methods) chronoamperometric current transients as shown in figure 1A-E on log-log curves. These transients show complex decays with at least two decay processes visible at long and short times.

We used equations (1) and (3) to calculate Q/M vs. C-rate, as plotted in each material as shown in figure 1F-J. We also measured rate-performance in a more traditional manner using GCD measurements (figure 1F-J). For those data sets, C-rate was calculated in the usual way, using equation (2) (see figure 1 F-J for theoretical capacities used). In addition, we used equations (5) and (6) to calculate Q/M vs. R curves as shown in figure 1 L-O. In addition, for comparison, we show capacity-rate data (open squares) measured using the traditional GCD technique (methods) with rate calculated using equation (4).

There are a number of points of interest in these data sets. The first point to note is that, in addition to the expected behaviour where we see a constant capacity at low rate followed by a fall off at higher rate, all capacity-rate curves extracted from CA show more complicated behaviour at the highest rates. For each material, for rates above C-rate~3C or R~50 h$^{-1}$, the capacity appears to level off somewhat before again falling away at higher rates. To our knowledge, such behaviour has not been reported before, possibly because such high rates are not generally explored in GCD measurements for electrodes as thick as we have used here. Interestingly, where we have performed GCD measurements at high rates (i.e. graphite and NCA) we see indications of this new behaviour (figure 1 G, H, L, M).

Secondly, we note that while Q/M vs. R curves measured using GCD and CA match extremely well, the agreement is somewhat poorer for the Q/M vs. C-rate curves. We attribute this to the fact that for the CA-derived curves, C-rate is normalised using the actual, low-rate capacity whereas for the GCD curves, C-rate is normalised (as usual) using the theoretical capacity. We would only expect good agreement for near-ideal electrodes with low-rate capacity approaching its theoretical value.

In addition, it is worth noting that for the Q/M vs. R curves, the agreement between CA and GCD data at low rate is extremely good. This vindicates our assertion that equation (6) can be used to obtain absolute capacities. Finally, we note that in the Q/M vs. R curves, the high rate decays for both normal and newly observed features are well-defined power laws. This is what would be expected in analogy with supercapacitors.[16] However, such well-defined power laws are not observed when capacity is plotted versus C-rate.

*Deriving a simple equation to fit capacity-rate data*

Equations (1), (3), (5) and (6) are extremely useful, not only because they can be used to transform current transients to capacity-rate data but because it is also possible to use them to generate simple equations to describe the dependence of capacity on rate. While simulations based on porous electrode theory can predict capacity-rate dependence,[17-18] it would be advantageous to have access to a simple equation which can be used to fit capacity rate data, outputting parameters which can be used to quantify rate-performance. A number of such equations have previously been proposed to describe the dependence of electrode capacity on charge discharge rate (see below for examples).[2, 7, 19-22] However, such equations tend to be empirical or semi-empirical. Here we use the concepts described above to derive a simple equation with the correct functional form to fit capacity rate data.

Here, we use the simplest possible circuit model for a battery electrode, treating it as a series R-C circuit.[23] In such a circuit, the capacitor charges gradually with a characteristic time, $\tau$, also known as the RC time constant. It is well known that for such a system the current response to a voltage step (magnitude V) is given by:

$$I = \frac{V}{R_s} e^{-t/\tau} \tag{7}$$

where $R_S$ is the series resistance and $\tau = R_s C$ where C is the capacitance. Then

$$Q/M = \int_0^t (I/M)dt = -\tau \frac{V}{R_s M}\left[e^{-t/\tau} - 1\right] \tag{8}$$

and

$$R = \frac{I/M}{\int_0^t (I/M)dt} = \frac{e^{-t/\tau}}{\tau\left[1 - e^{-t/\tau}\right]} \tag{9}$$

These equations can be combined by eliminating $e^{-t/\tau}$. Then, applying $\tau = R_sC$ leads to

$$\frac{Q}{M} = \frac{CV/M}{1 + R\tau} \tag{10}$$

In such a circuit, the charge on the capacitor at very long times is given by $Q_0 = CV$ such that:

$$\frac{Q}{M} = \frac{Q_0/M}{1 + R\tau} \tag{11}$$

However, long times mean very small currents which, because of the definition of R, signify very low rates. Thus, $Q_0/M$ represents the specific charge stored at low rate which is equivalent to the low-rate specific capacity, $Q_M$, allowing us to write:

$$\frac{Q}{M} = \frac{Q_M}{1 + R\tau} \tag{12}$$

This equation describes the dependence of specific capacity, Q/M, on rate R (as defined above) for a purely RC-limited battery electrode. This equation is consistent with Q/M being constant at low-rate (equal to $Q_M$). However, for higher rates the capacity decays inversely with rate: $(Q/M)_{high-rate} \approx Q_M / R\tau$. In this model, the rate-performance is controlled solely by $\tau$ which is a measure of the rate, $R_T$, at which the capacity starts to decay (defining $R_T$ as the rate where the low-rate constant capacity and high-rate approximation crossover yields, $R_T = 1/\tau$).

This model is extremely simple and does not cover the majority of cases where rate-performance is diffusion limited. However, we recently showed that an equation such as this can be empirically modified to allow the fitting of data for electrodes which are electrically-limited (i.e. limited by the RC charging time), diffusion limited, or limited by a combination of both. To see this, we note that, at high rate, diffusion limited battery electrodes tend to display $(Q/M)_{high-rate} \propto \tau^{-1/2}$, distinct from the $(Q/M)_{high-rate} \propto \tau^{-1}$ behaviour described above for electrically-limited electrodes.[7] Electrodes limited by a combination of RC and diffusion effects tend to display $(Q/M)_{high-rate} \propto \tau^{-n}$, where usually $0.5 \leq n \leq 1$.[7] These variations in high rate behaviour represent the main difference between the behaviour of real battery electrodes and the idealised case described by equation (12).

This possibility for variations in high rate behaviour where n may be different from 1, can be incorporated into equation (12), simply by adding an empirical exponent, n, which will allow the equation to match all capacity-rate data. We have previously shown that such an empirical modification to an equation originally derived to describe supercapacitors is very effective for describing batteries.[7] In addition, we have added a factor of 2, a modification which does not change the functional form of the equation but rather modifies the value of the time constant.

$$\frac{Q}{M} = \frac{Q_M}{1+2(R\tau)^n} \qquad (13)$$

As shown in the SI and below, this factor of 2 means the τ-value in equation (13) is now consistent with that in the equation proposed by Tian et al.[7] As shown in the SI, adding this factor means that when Tian's equation and equation (13) are used to fit the same data set, the fit values outputted are similar. This modification is useful because Tian et al have reported a large dataset of τ-values so it is important to be consistent with their results.[7]

Once these modifications have been made, the parameter, τ, in equation (13) no longer represents the RC time constant but a more general characteristic time associated with charge discharge. In addition, τ is now related to $R_T$ by $R_T = (1/2)^{1/n}/\tau$. As described in Tian et al, this characteristic time can then be linked to mechanistic models.[7]

Alternatively, we can perform a very similar analysis to that above to find an expression for Q/M as a function of C-rate (which we write as $R_C$). To do this, we use equations (7) to perform the integrations in equations (1) and (8) obtaining:

$$\frac{Q}{M} = Q_M\left[1-\tau R_C\right] \qquad (14)$$

We note that such a linear decay of Q/M with C-rate has previously been proposed by Doyle et al for electrically limited battery electrodes.[24]

As above, this equation can be generalised by adding a factor of 2 and an exponent:

$$\frac{Q}{M} = Q_M\left[1-2(\tau R_C)^n\right] \qquad (15)$$

Including this factor of 2 means that the time constant in equation (15) is now consistent with that in Tian's equation[7] as well as the τ-value equation (13). As before, once these modifications have been made, the parameter, τ, in equation (15) no longer represents the RC time constant but a more general characteristic time associated with charging and discharging.

*Comparison with other capacity versus rate equations*

A number of empirical equations relating capacity to rate have been proposed previously. Here we reproduce some of those equations, modifying the parameters slightly to match the notation used here. In addition, we add a factor of 0.5 to one of the equations (that of Heubner et al.) which was not present in its originally proposed form. Again, this factor does not change the functional form of the equation but, modifies the value of the time constant outputted when fitting a given curve. As shown in the SI and below, this factor allows the time constant to be consistent with both the Tian's equation and equation (13).

$$Q/M = Q_M\left[1-(R\tau)^n\left(1-e^{-(R\tau)^{-n}}\right)\right] \qquad \text{(16, Tian et al)}^{7}$$

$$Q/M = Q_M\left(1-\exp\left[-0.5\left(R_C\tau\right)^{-n}\right]\right) \qquad \text{(17, Heubner et al)}^{21}$$

$$Q/M = Q_M\exp\left[-\left(R_C\tau\right)^{n}\right] \qquad \text{(18, Wong et al)}^{2}$$

The Tian et al. equation is from our previous work and was expressly proposed in terms of rate R.[7] However, the Heubner and Wong equations were originally expressed in terms of C-rate, $R_C$. However, it is worth considering whether equations such as these are best expressed in terms of R or C-rate.

For comparison purposes, we plot equations (13) and (15-18) on figure 2A-B (in terms of a generic rate). We find that the equations (16) and (17) (those of Tian and Heubner) have a form which is very similar to that of equation (13) (figure 2A). These three equations are almost identical: constant at low rate and decaying as a power-law at high rate, differing only slightly from each other in the turnover region (i.e. for rates close to $R_T$). Because both equation (13) and (16) (Tian's equation) were expressly derived to describe Q/M vs. R data, we propose that Heubner's equation is more naturally suited to fitting data plotted versus R than C-rate as originally proposed. As a result, we suggest the $R_C$ in equation (17) should be replaced by R.

Equations (15) and (18) (Wong's equation) are also similar to each other but clearly different in form to the other three equations (figure 2B). Both decay rapidly above $R_T$ and neither show power-law decays at high rate. We note that Wong's equation was proposed to describe capacity versus C-rate data while equation (15) was explicitly derived in terms of C-rate.

We can clarify the behaviour of these equations by reproducing the Q/M vs. rate data for graphite as obtained from CA, plotted both versus R but also plotted versus C-rate in figure 2C. For clarity, we have only included the first decay at lower rates, ignoring the second feature at very high rates. Both curves are similar at low rates beginning to deviate only at rates above $R_T$. At high rate the Q/M vs. R data falls off as a power law while the Q/M vs. C-rate falls off much more rapidly and, when examined closely, is not power-law like. The form of this experimental data strongly suggests that the equations plotted in figure 2A effectively describe Q/M vs. R data while those plotted in figure 2B are appropriate for fitting Q/M vs. C-rate data.

*Fitting CA data*

The analysis described above suggests that equation (13), (16) or (17) should be used for fitting Q/M vs. R data while equations (15) or (18) should be used to fit Q/M vs. C-rate ($C_R$) data. As shown in the SI (fig S1), when fitting Q/M vs. R data, we find that all three equations give good fits yielding very similar fit parameters. Equation (13) appears to give marginally better fits to the data. Similarly, when fitting Q/M vs. C-rate data, we find that both equations give good fits (fig S2). However, the fit parameters are very different from each other, although those obtained using equation (15) are in reasonable agreement to the fit parameters contained from the Q/M vs. R fits. Thus, we limit ourselves to using equation (13) to fit Q/M vs. R data and equation (15) to fit Q/M vs. C-rate data. However, which equation one uses is to some degree a matter of taste.

We next used equation (13) to fit Q/M vs. R data for all five materials, generated both by GCD and CA, as shown in figure 3. In addition, we have used equation (15) to fit Q/M vs. $C_R$ data for all materials, generated both by GCD and CA (figure 4). In both cases, we have fit only the low rate portion of the CA curves corresponding to the standard, commonly observed falloff of capacity with rate.

The fits of the Q/M versus rate data shown in figure 3 and 4 are all very good. However, we note that such fitting required limiting the fitting range to avoid the start of the second decay feature at higher rates. In both figures 3 and 4, we have extended the range of rates shown to include the start of this second decay to illustrate its presence. It can clearly be seen as the region of high rate where the fit deviates from the data. The fit parameters extracted from fitting the data in figures 3 and 4 are shown in tables 1 and 2 respectively.

The fit parameters associated with the Q/M vs. R data, measured by both GCD and CA agree with each other very well. The fractional deviation (e.g. $(\tau_{GCD}-\tau_{CA})/\tau_{GCD}$) between equivalent fit parameters for data measured by GCD and CA ranged from <1% to 23% with a root-mean-square (RMS) fractional deviation calculated over all parameters of only 10%.

Similarly, the fit parameters associated with the CA data, plotted versus both R and C-rate agree with each other reasonably well with fractional deviations also varied between <1% and 23% (with a single outlier at 42%) and an RMS fractional deviation of only 15%.

However, the agreement between fit parameters associated with the Q/M vs. C-rate data, measured by GCD and CA was much poorer. For these data sets, the individual fractional deviations were in the range of 2% to 170% and had a RMS fractional deviation of 72%.

We can summarise this result by saying that the equivalent fit parameters extracted from the Q/M vs. R data obtained from GCD and CA and the parameters from the Q/M vs. C-rate data obtained from CA were all consistent with each other while the fit parameters from the Q/M vs. C-rate data obtained from GCD deviated significantly from the rest. The variation in fit parameters is shown graphically in figure S3.

We believe the poor agreement between Q/M vs. C-rate data obtained from GCD and the rest stems from that fact that for this data set, C-rate is calculated using the theoretical capacity which is an arbitrary value which is not necessarily closely correlated with low-rate capacity for all electrodes. For this reason, we believe that when quantitative rate analysis is required it is best to fit Q/M vs. R data obtained either from GCD or CA measurements. However, if C-rate data is required, we believe it is best to obtain it from CA as described above.

*The high rate region*

Thus far, we have focused on the low-rate region of the capacity-rate curves, a region which is consistent with almost all reported capacity-rate data sets. However, as shown in figure 1, there is also a second region at high rate where an additional mechanism appears to be contributing to the decay of capacity with rate. To examine this high-rate decay region, we performed cyclic voltammetry (CV) measurements on an NCA-based electrode over a wide range of scan rates.

Shown in figure 5A is the Q/M v R curve, measured by CA, for such an electrode. On the top axis, we show the equivalent CV scan rates, ν, estimated from $v = R\Delta V$ with $\Delta$V=1.3V. We have divided this curve into four zones, delineated by the vertical dashed lines. Zones 1 and 2 represent the low-rate region normally observed in capacity-rate experiments with zone 1 representing the constant capacity regime and zone 2 representing the initial capacity decay. However, like those curves in figure 1K-O, the curve in figure 5A contains a second decay at high rate with zone 3 representing the first stage of this decay-process and zone 4 representing the final stage in this process where capacity decays roughly inversely with R.

Shown in figure 5B-E are CV curves arranged by zone associated with their scan rates. The lowest rate CV curve is shown in figure 5B and is associated with zone 1. This curve is typical of NCA at low rate[25] and contains oxidation and reduction features at 3.85 and 3.6 V and fully consistent with known intercalation reactions for NCA. As the scan rate increases into zone two, the current increases slowly while at the same time, the curves get more and more compressed as is typically observed at higher scan rates.[26] This feature is consistent with an electrode behaving as a resistive pseudocapacitor, without redox activity linked to intercalation reactions. However, as shown in figure 5D, by the time the scan rate increases into zone 3 (where the second high-rate decay becomes important, see figure 5A), the CV curves have started to transition from high-rate redox curves to resistive CV curves which enclose only very small areas and are almost completely linear in potential. For scan rates in zone 4 (figure 5E), the curves are almost completely resistive.

We can understand this scan rate dependence in more detail by plotting the current at various potentials versus scan rate, ν, in figure 5E. In each case, the current initially increases roughly as $\nu^{1/2}$, before saturating at high scan rate. This $\nu^{1/2}$ behaviour is indicative of diffusive limitations within the electrolyte and is usually described via the Randles–Sevcik equation:[27]

$$I_P = 0.4463nFAC\left(\frac{nFDv}{RT}\right)^{1/2} \tag{19}$$

where $I_P$ is the peak current (A), n=1 is the number of electrons transferred in the redox event, A is the electrode area ($cm^2$), D is the diffusion coefficient ($cm^2/s$), C is the ionic concentration ($mol/cm^3$) and other parameters have their usual meaning (and SI units). Assuming $D=10^{-6}$ $cm^2/s$, an approximate value for Li ions diffusing in electrolyte,[28-29] we plot this equation on figure 5F (grey line), finding very good agreement with the low-scan-rate data.

However, we find this diffusion limitation to hold only in zones 1 and 2 (i.e. where the Q/M vs/ R curve displays normal behaviour). Once, the scan rate enters zone 3, the current begins to saturate, becoming limited by resistive effects as described above, and devoid of anodic and cathodic current peaks. This implies that, in this case, the first decay in the low rate portion of the Q/M v R curve is predominately limited by diffusion effects while the second decay at high rate is limited by electrical effects. In this latter case, high rate behaviour that appears as a capacitance on a severely resistive background is described by remnant double layer capacitance, and the transition to intercalation reactions cannot proceed owing to limited electron availability at the electrode surface at high rates.

One possible explanation for this would be to note that the CA data is extracted from current transients such as those shown in figure 1A-E. Because R is directly proportional to the current (which decays with time), high rates are associated with short times and vice versa. This means the high-rate portion of the Q/M v R curve is equivalent to the early-time portion of the current transient. At very short times, ions do not have significant time to move, meaning that the current response is dominated by electron motion in the electrode. This would imply that the very early stage of the current transient is predominately electrically limited. Only at later times would other factors such as ion diffusion start to become significant. Such slower processes would explain the second decay in the current transients. In terms of Q/M vs R curves, this would mean that while the second, high-R decay is controlled by only electrical properties of the electrode, the first low-R decay is controlled by all rate-limiting effects with diffusion limitations being a major contributor. Ultimately, this implies that the first, low-rate decay in the Q/M vs. R curves is most relevant as it contains information about the overall set of factors limiting rate-performance.

*Fitting both components of Q/M v rate curves*

If we treat both features in the Q/M vs. R curves separately, then we can empirically fit the entire curve as the sum of two capacity decays, each of which is described by equation 13:

$$\frac{Q}{M} = \frac{Q_{M,1}}{1+2(R\tau_1)^{n_1}} + \frac{Q_{M,2}}{1+2(R\tau_2)^{n_2}} \tag{20}$$

where the subscripts 1 and 2 refer to the low- and high-rate decays respectively. As shown in figure 6 for the Q/M vs. R data, this equation matches the data extremely well with residuals of less than 10% over all materials for all rates. The fit parameters are given in table 3 for both high- and low-rate components. In all cases, the fit parameters for the low-rate component match well to those found by fitting only the low-rate range in figure 3 F-J (table 1).

The ability to extract parameters which quantitatively describe the high-rate decay (in particularly $\tau_2$) allow us to explore this feature in more detail. To do this, we prepared NCA electrodes with a range of thicknesses from 25-120 μm. We then measured the rate-performance by CA for each electrode with examples shown in figure 7A. All Q/M vs. R curves were fit using equation 20 (see table S3). Over all electrodes, $Q_{M,1}$, $Q_{M,2}$, $n_1$ and $n_2$ were roughly thickness invariant with averages of 185 mAh/g, 4.1 mAh/g, 0.96 and 1.05. However, both $\tau_1$ and $\tau_2$ both varied with electrode thickness as shown in figure 7B. While both time constants increase with $L_E$, $\tau_1$ is roughly ×100 larger than $\tau_2$. In addition, the $L_E$-dependence looks slightly different for each time constant.

We believe that, for CA-obtained rate data, the low-rate time constant, $\tau_1$, is equivalent to the time constant ($\tau$) obtained by fitting equation 16 to capacity-rate data obtained by GCD. Recently, we proposed a simple mechanistic model,[7] which expressed $\tau$ as the sum of contributions due to the RC time constant of the electrode, various diffusion times and the electrochemical reaction time.[7] Applying this model to $\tau_1$ gives:

$$\tau_1 = L_E^2\left[\frac{C_{V,eff}}{2\sigma_E}+\frac{C_{V,eff}}{2\sigma_{BL}P_E^{3/2}}+\frac{1}{D_{BL}P_E^{3/2}}\right]+L_E\left[\frac{L_S C_{V,eff}}{\sigma_{BL}P_S^{3/2}}\right]+\left[\frac{L_S^2}{D_{BL}P_S^{3/2}}+\frac{L_{AM}^2}{D_{AM}}+t_c\right] \quad (21a)$$

where $C_{V,eff}$ is the effective volumetric capacitance of the electrode (F $m^{-3}$), $\sigma_E$ is the out-of-plane electrical conductivity of the electrode material (S $m^{-1}$), $\sigma_{BL}$ is the overall (anion and cation) conductivity of the bulk electrolyte (S $m^{-1}$), $D_{BL}$ is the diffusion coefficient of Li ions in the bulk electrolyte ($m^2$ $s^{-1}$), $P_E$ and $P_S$ are the porosities of the electrode and separator respectively while $L_S$ is the separator thickness (m). In addition, $L_{AM}$ is a measure of the size of the active particles (m); $D_{AM}$ is the solid state Li ion diffusion coefficient within the particles ($m^2$/s) while $t_c$ is a measure of the timescale associated with the electrochemical reaction once electron and ion combine at the active particle (s). This equation has been shown to accurately describe a wide range of experimental data and makes predictions which are consistent with observations.[7] For brevity, it is easier to express this equation in our case as follows:

$$\tau_1 = aL_E^2 + bL_E + c \quad (21b)$$

Fitting equation 21b to the data for $\tau_1$ vs. $L_E$ in figure 7B yields an excellent fit with parameters a=7.3×$10^{10}$ s/$m^2$, b=5.7×$10^5$ s/m, and c=101 s. We note that the values of and b are in line with those measured for other systems[7] while the value of c is relatively high compared to other systems,[7] indicating solid state diffusion is important in this electrode.

We have argued above the high-rate feature (time constant $\tau_2$) is due to rate-limiting factors associated only with electron motion because all other process are too slow to contribute in the very early states of the transient. Because equation 21a contains all rate-limiting factors, we argue that one or more of its terms should describe $\tau_2$. Within equation 21a, only the first term is due to the transport of charge within the electrode. This implies that only this term can contribute to $\tau_2$. Then, using an empirical observation relating the electrode capacitance to its volumetric capacity ($C_{V,eff}/Q_V = 28$ F/mAh),[7] and taking $Q_V = \rho_E Q_M$, where $\rho_E$=3.2 g/$cm^3$ is the measured mean electrode density, we obtain:

$$\tau_2 = \frac{14\rho_E Q_M}{\sigma_E} L_E^2 \quad (21c)$$

where $Q_M$ should be expressed in mAh/kg. We have fit this equation to the data in figure 7B finding good agreement. Combing the measured density and out-of-plane conductivity of the electrode (3200 kg/$m^3$ and 0.27 S/m) with the fit parameter yielded a value of $Q_M$=0.29 mAh/g. This value is much smaller than the low-rate specific capacity of the NCA electrodes which was approximately 190 mAh/g (the fits to equation 20 gave a mean value of $Q_{M,1}$ of 185 mAh/g for NCA electrodes of arrange of thicknesses). However, this value is close to the specific capacitance associated with the high-rate component of the fits to the Q/M vs. R curves for NCA electrodes of different thicknesses. Averaging over fit values for 11 electrodes of different thickness gave $\langle Q_{M,2} \rangle$ = 4.1±0.9 mAh/g, in very good agreement with the value above.

This implies that $\tau_2$ is related to $Q_{M,2}$, via

$$\tau_2 = \frac{14\rho_E Q_{M,2}}{\sigma_E} L_E^2 \qquad (21d)$$

If this is the case, then the scatter in $\tau_2$ observed in figure 7B might be related to scatter in $Q_{M,2}$. To check this we plotted $\tau_2 / L_E^2$ vs. $\rho_E Q_{M,2}$ in figure 7C with each data point calculated from the fit parameters for each of the electrode thicknesses described above. This data does indeed show a linear relationship consistent with equation 21d. To confirm, this we plotted the prediction of equation 21d (solid line) using the measured out of plane conductivity for an NCA/SWNT electrode (0.27 S/m). This line agrees very well with the data supporting the electrical origin of the high rate feature. We note that when applying equation 21a to $\tau_1$ in conjunction with $C_{V,eff} / \rho_E Q_M = 28$ F/mAh, the relevant value of $Q_M$ is that representing the total low-rate capacity.[7] With reference to equation 20 the total low-rate capacity is $Q_{M,1} + Q_{M,2}$. However, because $Q_{M,1} >> Q_{M,2}$, it would be reasonably accurate to use $Q_{M,1}$ in equation 21a when analysing data for $\tau_1$.

It is not clear why the capacity in equation 21c should be associated with that of the high-rate term in equation 20, or indeed what determines the value of $Q_{M,2}$, although it may just be the capacity associated with double layer charge storage. In any case, the data described here strongly suggests that high-rate feature in the Q/M vs. R curves derived from CA is electrical in nature and might be used to obtain valuable information about electrode operation.

## CONCLUSIONS

In summary, we have demonstrated the use of chronoamperometry as a method for obtaining capacity vs. rate data for a number of common electrode materials. In addition to converting

CA current transients to capacity versus C-rate data, we demonstrate how to generate capacity versus charge/discharge rate, R. Curves of capacity versus either C-rate or R have similarities but also well-defined differences. Such curves can be obtained reasonably quickly and tend to have hundreds or even thousands of data points down to C-rates as low as 0.01C. We note that it would be extremely time consuming to obtain such data using standard methods. In addition to the normally observed fall-off of capacity with rate, we see an additional capacity decay feature at high rate which we believe is previously un-reported. We find very good agreement between CA-based measurements and traditional galvanostatic charge-discharge data, even in the high-rate regime where the new feature is observed.

In order to perform quantitative analysis on the rate data obtained by CA, we have applied the equations used to transform current transients to capacity-rate data to generate equations for specific capacity as a function of both C-rate and charge/discharge rate, R. These equations fit data well and mirror the differences in data between C-rate and R plots. This allows us to identify sets of models from the literature which are appropriate to fitting capacity v C-rate or capacity v R data. Finally, we have investigated the differences between the normal, low rate region of the capacity-rate curves and the new feature at high rate. In an NCA-based electrode, we see the low-rate region of the capacity-rate curve to be predominately diffusion limited whereas the high-rate region appears to be electrically limited. We attribute this to the fact that the high-rate region is associated with the short-time portion of the transient. At very short time, ions have not had enough time to move meaning the only contributions to the current are electronic. This hypothesis explains the observed quadratic scaling of high-rate time constant with electrode thickness and allows this data to be used to estimate the out-of-plane electrode conductivity.

## METHODS

The NCM, Graphite and LTO electrodes were commercial electrodes (Customcells, Germany). The mass loadings and compositions were $LiNi_{0.33}Mn_{0.33}Co_{0.33}O_2$ (NMC111, 34 mg/cm$^2$ total loading, thickness 130 μm, 86 wt% NMC111 + 14% additives), graphite (14 mg/cm$^2$ total loading, thickness 95 μm, 86 wt% Graphite + 14% additives), and $Li_4Ti_5O_{12}$ (LTO, 32 mg/cm$^2$ total loading, thickness 184 μm, 84 wt% LTO + 16% additives). The other two were fabricated from NCA and a mixture of micron-Si/Graphite both incorporating carbon nanotubes (CNT, Tuball, OCSiAl) as a conductive additive. Here we employed $LiNi_{0.815}Co_{0.15}Al_{0.035}O_2$ (NCA, MTI, 20 mg/cm$^2$ total loading, thickness 65 μm,) as an active material as well as micro-sized silicon (1~3 µm, US Research Nanomaterials) mixed with graphite powder (C-NERGY SFG 6L RAPHITE) (labeled as µSiGr, 1.9 mg/cm$^2$ total loading, thickness 19 µm). The compositions of these electrodes were Si/Graphite/CNT (60:33:7 by wt) and NCA/PVDF/CNT (94:5:1 by wt). The AM powers were directly mixed with the CNT solution and ground by a mortar and pestle to obtain a uniform slurry. Then the slurry was cast onto either Al or Cu foil using a doctor blade. Then the slurry cast electrodes were slowly dried at 40 °C for 2 hours and followed by vacuum drying at 100 °C for 12 hours to remove residual water. The areal capacity for all the electrodes was approximately 4 mAh/cm$^2$. In addition, we made a set of NCA-based electrodes, all with composition NCA/PVDF/CNT (94.5:5:0.5 by wt), but for a range of mass loadings from 8-38 mg/cm$^2$ (25-120 µm).

The electrochemical properties of the electrodes were measured in 2032-type coin cells (MTI Corp.) with a half-cell configuration. All coin cells were assembled in an Ar-filled glovebox (UNIlab Pro, Mbraun). The dried electrodes were cut into 12 mm diameter discs (area = 1.13 cm$^2$) and paired with Li metal discs (diameter= 16 mm). Celgard 2032 (thickness = 16 µm) was used as a separator. The electrolyte was 1.2 M $LiPF_6$ dissolved in EC/EMC (1:1 in v/v, BASF) with 10 wt% fluoroethylene carbonate (FEC).

Electrochemical tests were performed using a potentiostat (VMP3, Biologic) with voltage ranges of 0.005−1.2 V, 0.001−1.5 V, 1−2.5 V and 3−4.3 V for µSiGr, Graphite, LTO, and NCM/NCA electrodes, respectively. All the cells were firstly performed GCD at 1/20 C for 1 cycle and at 1/10 C for 5 cycles. After the capacities were stable (capacity change is <1%), and coulombic efficiency is >99%, the cells were charged to upper cut off potential at 1/10 C, and then the CA were performed at lower cut off potential. More details are given in table 3. While a wider voltage range with same lower cutoff would give a larger capacity, we chose a voltage range that maximized coulombic efficiency without significant electrolyte decomposition (CEI or SEI formation) from cycle to cycle. We note that in this work, the same voltage ranges were used for CA and GCD, to maximise the agreement between measurements.

**Acknowledgments:** The authors acknowledge the SFI-funded AMBER research centre (SFI/12/RC/2278) and Nokia for support. JNC thanks Science Foundation Ireland (SFI, 11/PI/1087) and the Graphene Flagship (grant agreement n°785219) for funding. COD acknowledges SFI under grant nos 13/TIDA/E2761, 14/IA/2581, 15/TIDA/2893 and the SmartVista project, which has received funding from the European Union’s Horizon 2020 research and innovation programme under the grant agreement No. 825114.

**Supporting Information**. Brief description of how to make capacity-rate equations consistent with each other. Comparison of different capacity-rate equations fitting both GCD and CA data including fit parameters. Fit parameter table for thickness dependent study.

Figures

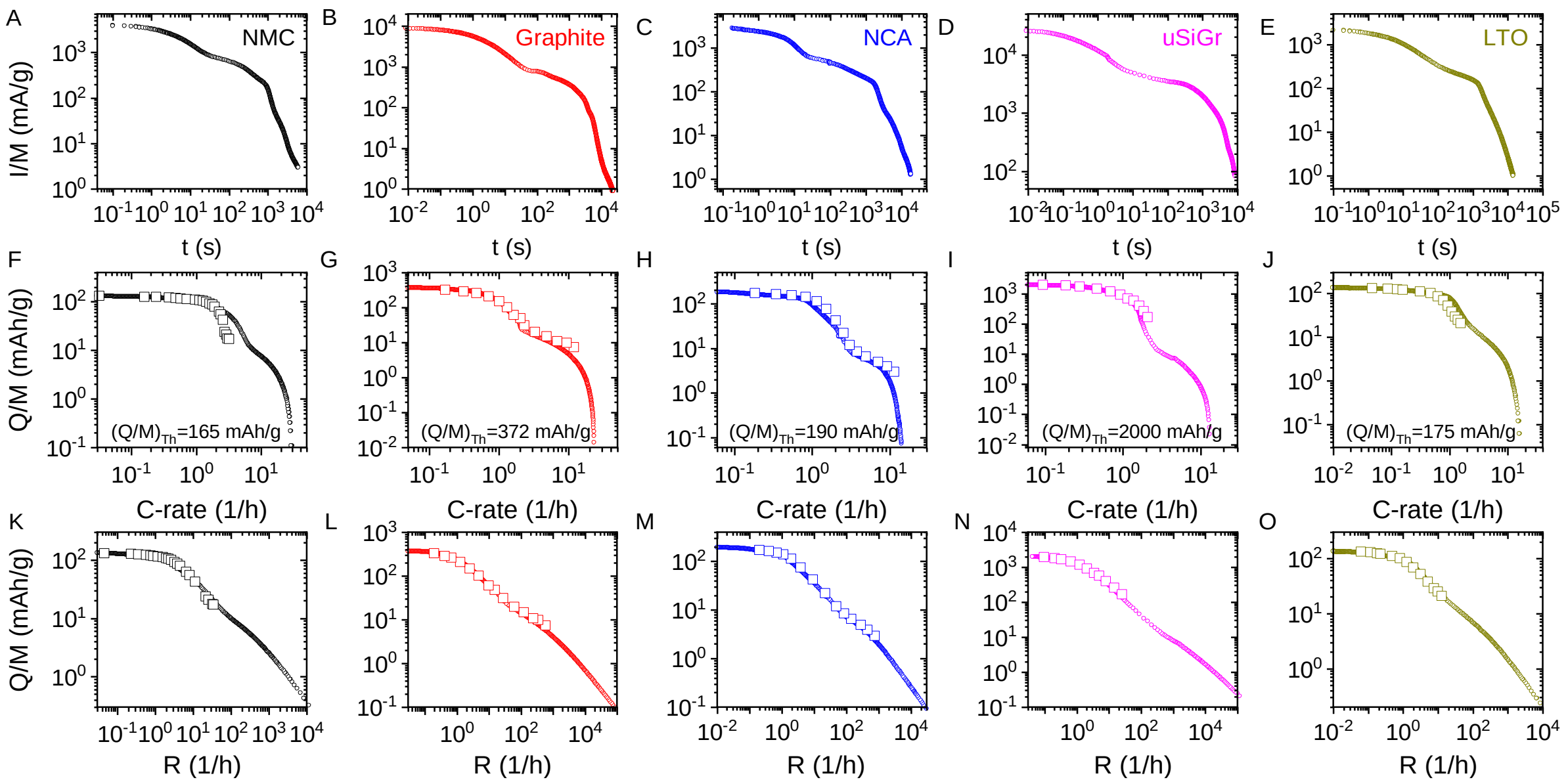

Figure 1: A-E) Current transients measured by chronoamperometry for five lithium storing materials, NMC111, graphite, NCA, μSiGr and LTO. F-P) Associated Q/M vs C-rate ($R_C$) curves (F-J) and Q/M vs R (K-O), both extracted from the current transients for the same materials (colour coding is the same in each row). In F-P, the equivalent data measured by galvanostatic charge discharge methods is plotted as the larger open squares. The text panels F-J shows the theoretical capacities used to calculate the C-rates for the GCD data. In all cases, capacities are normalised to the active electrode mass.

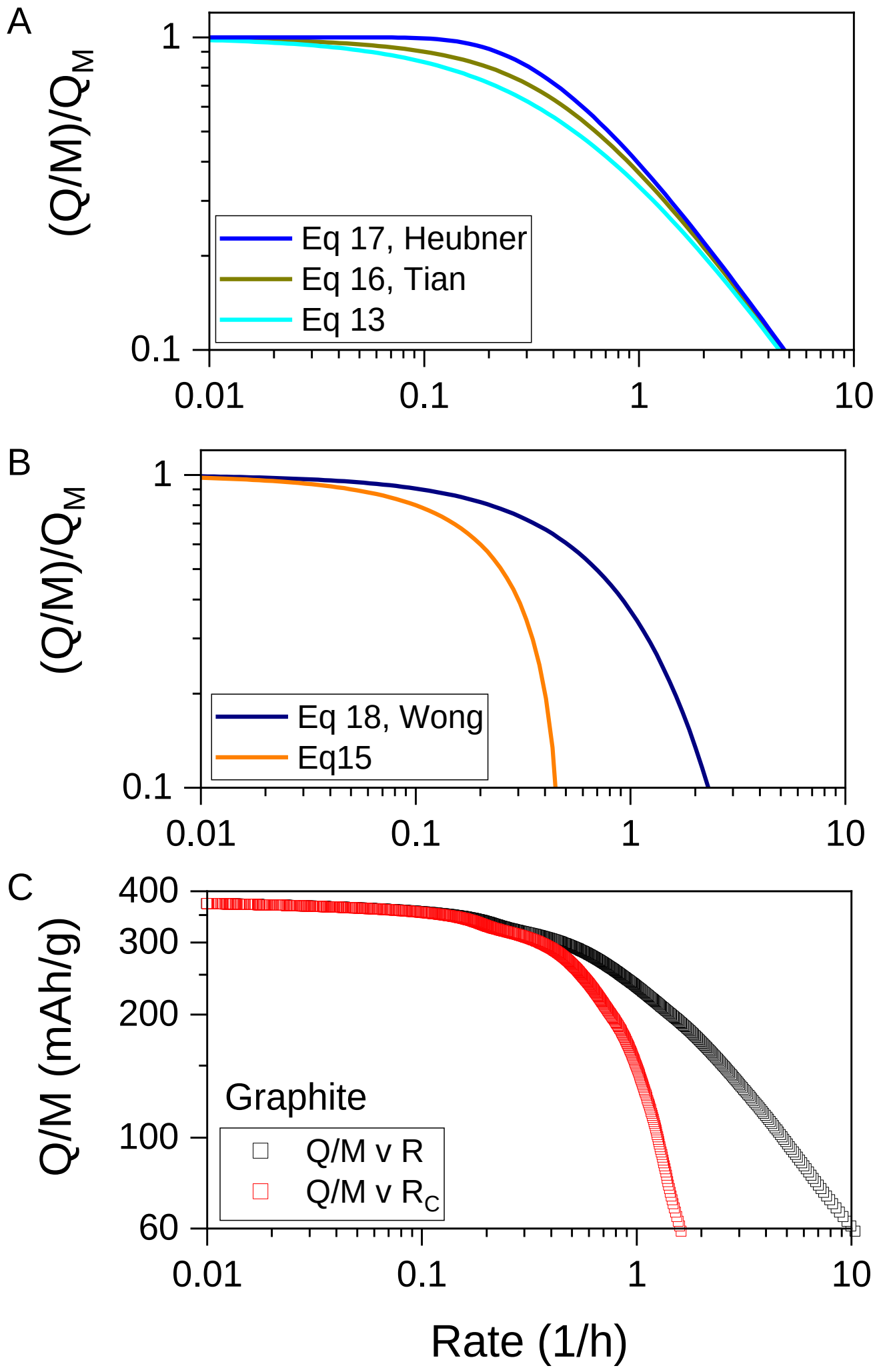

Figure 2: A-B) Comparison of equations (13) and (15) with three empirical capacity-rate equations taken from the literature. In figure 2A equation (13) is compared to Heubner's equation (eq 17) and Tian's equation (eq 16). In figure 2B equation (15) is compared to Wong's equation (eq 18). In all cases, the curves are plotted using $\tau$=1 and n=1. C) Capacity-rate data for a graphite electrode plotted as Q/M vs. R (black) and Q/M vs. C-rate (red).

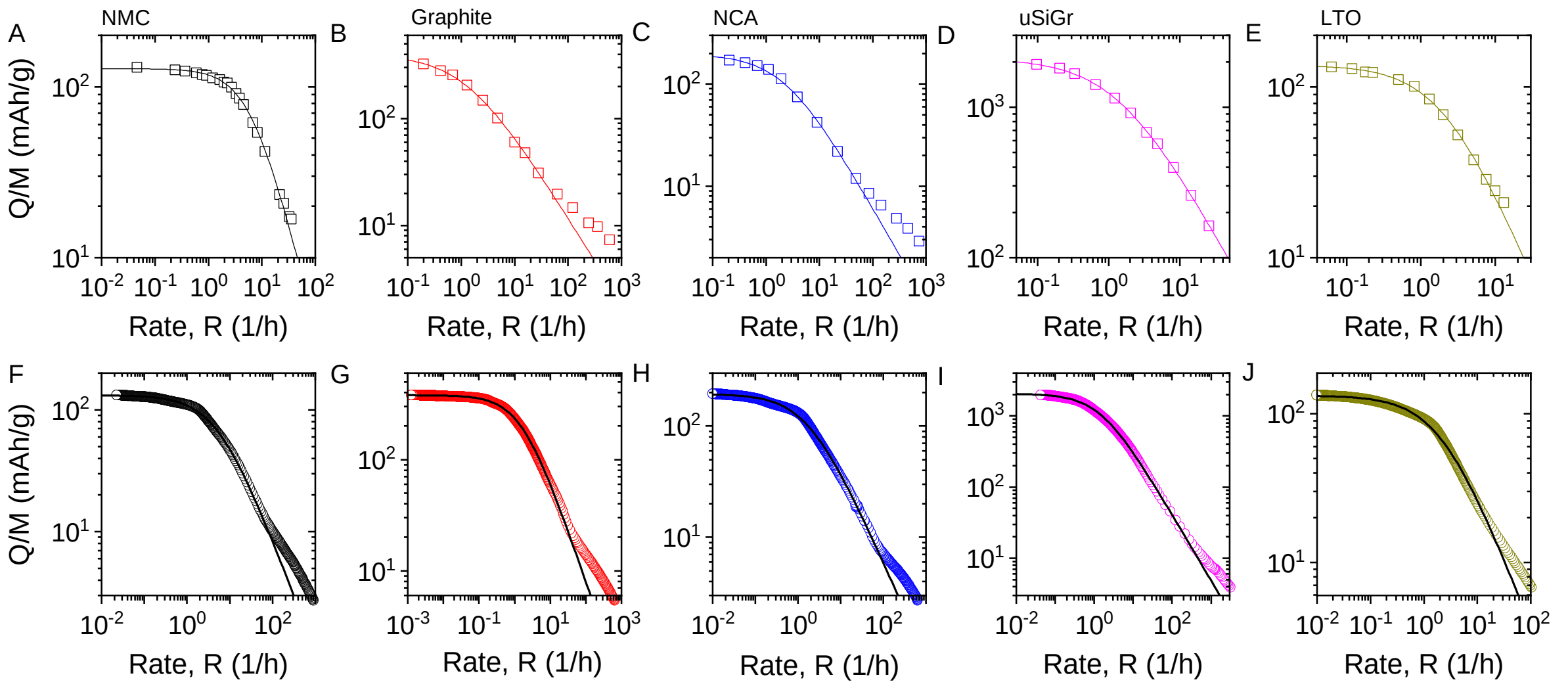

Figure 3: Q/M vs. R curves, measured by GCD (A-E) and chronoamperometry (F-J) for A,F) NMC111, B,G) graphite, C,H) NCA, D,I) uSi-Gr and E,J) LTO. Each curve is fitted to equation (13), with the fit limited to the low rate region.

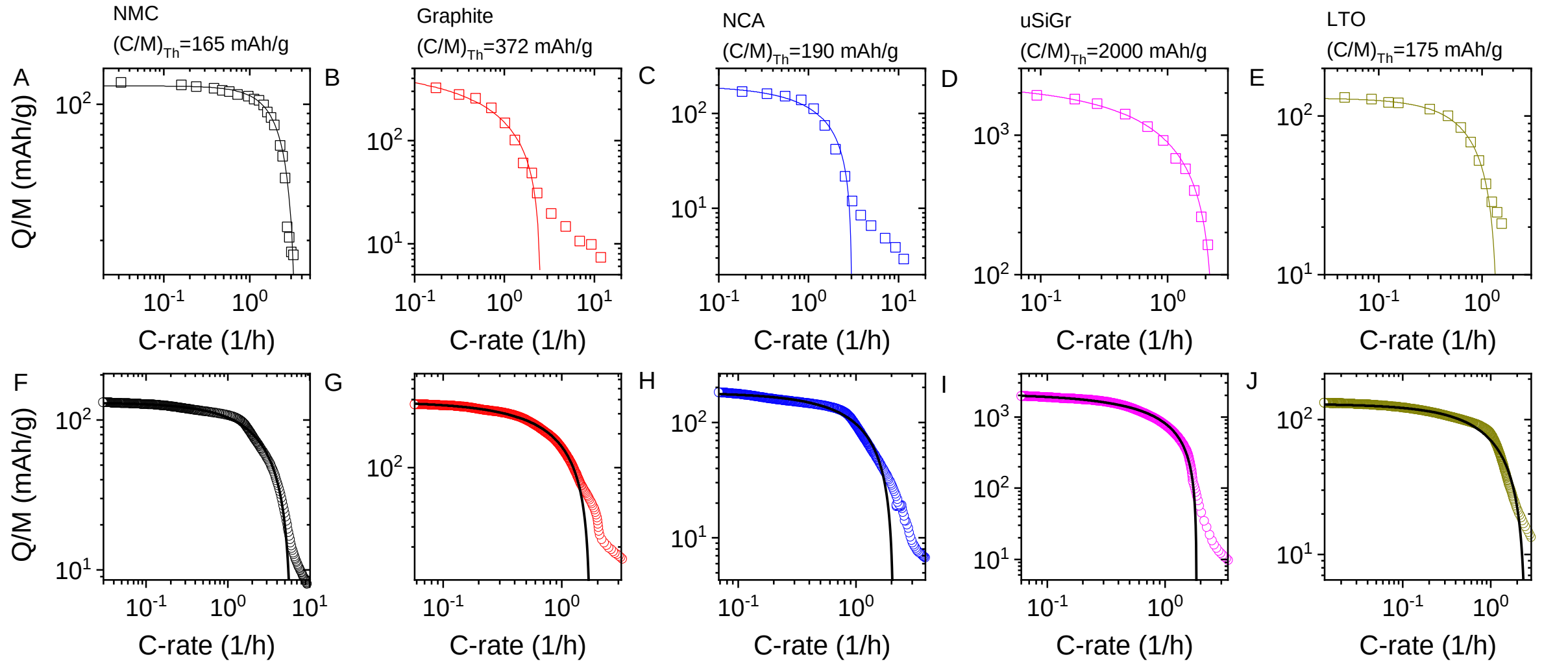

Figure 4: Q/M vs. C-rate ($R_C$) curves, measured by GCD (A-E) and chronoamperometry (F-J) for A,F) NMC111, B,G) graphite, C,H) NCA, D,I) uSi-Gr and E,J) LTO. Each curve is fitted to equation (15), with the fit limited to the low rate region.

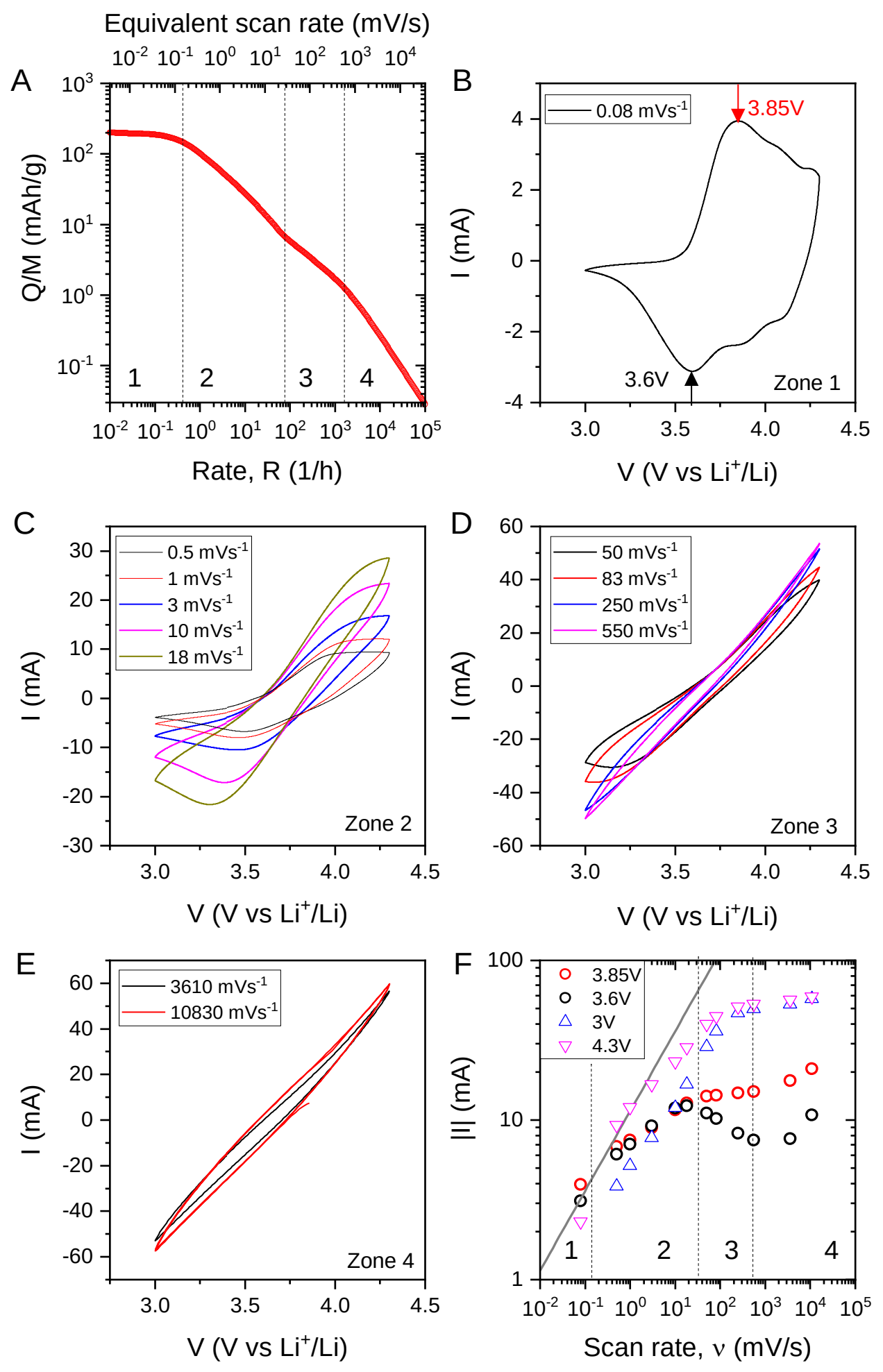

Figure 5: A) Example of a capacity vs. rate curve, obtained via CA, for an LiNiCoAlO2 (NCA) based electrode with M/A=32 mg/cm$^2$ and thickness of 105 μm. The equivalent scan rate for a CV curve is given in the top axis. The vertical dashed lines separate the graph into four zones. Zone 1 is the low rate region of constant capacity. Zone 2 is the typically observed region where the capacity falls at higher rates. Zone 3 represents a flattening of the capacity -rate curve at very high rates while zone 4 represents the ultimate decay of capacity at extremely high rates. B-E) Cyclic voltammetry curves collected at scan rates associated with zones 1 (B), 2 (C), 3 (D) and 4 (E). F) Current measured in CV experiment at four different potentials plotted as a function of scan rate. The vertical dashed lines delineate the zones mentioned above. The solid gray line represents a plot of the Randles–Sevcik equation assuming D=10$^{-6}$ cm$^2$/s.

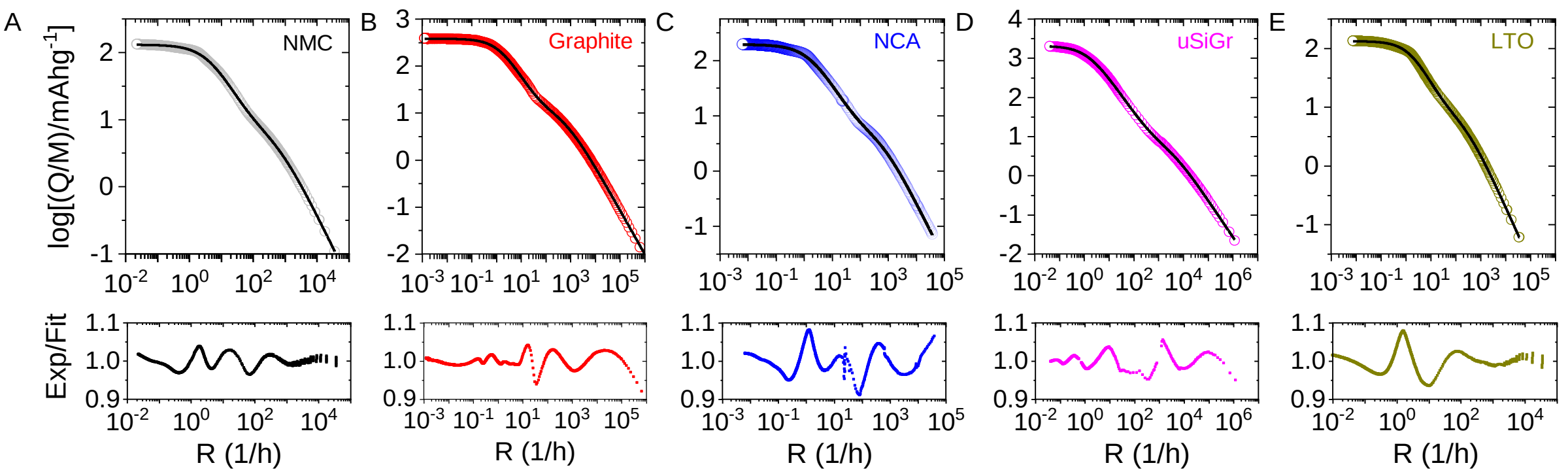

Figure 6: Capacity-rate curves, measured by chronoamperometry fitted using equation 20 for A) NMC111, B) graphite, C) NCA, D) μSi-Gr and E) LTO. The lower panels show the fitting residuals, expressed as experimental data divided by fit.

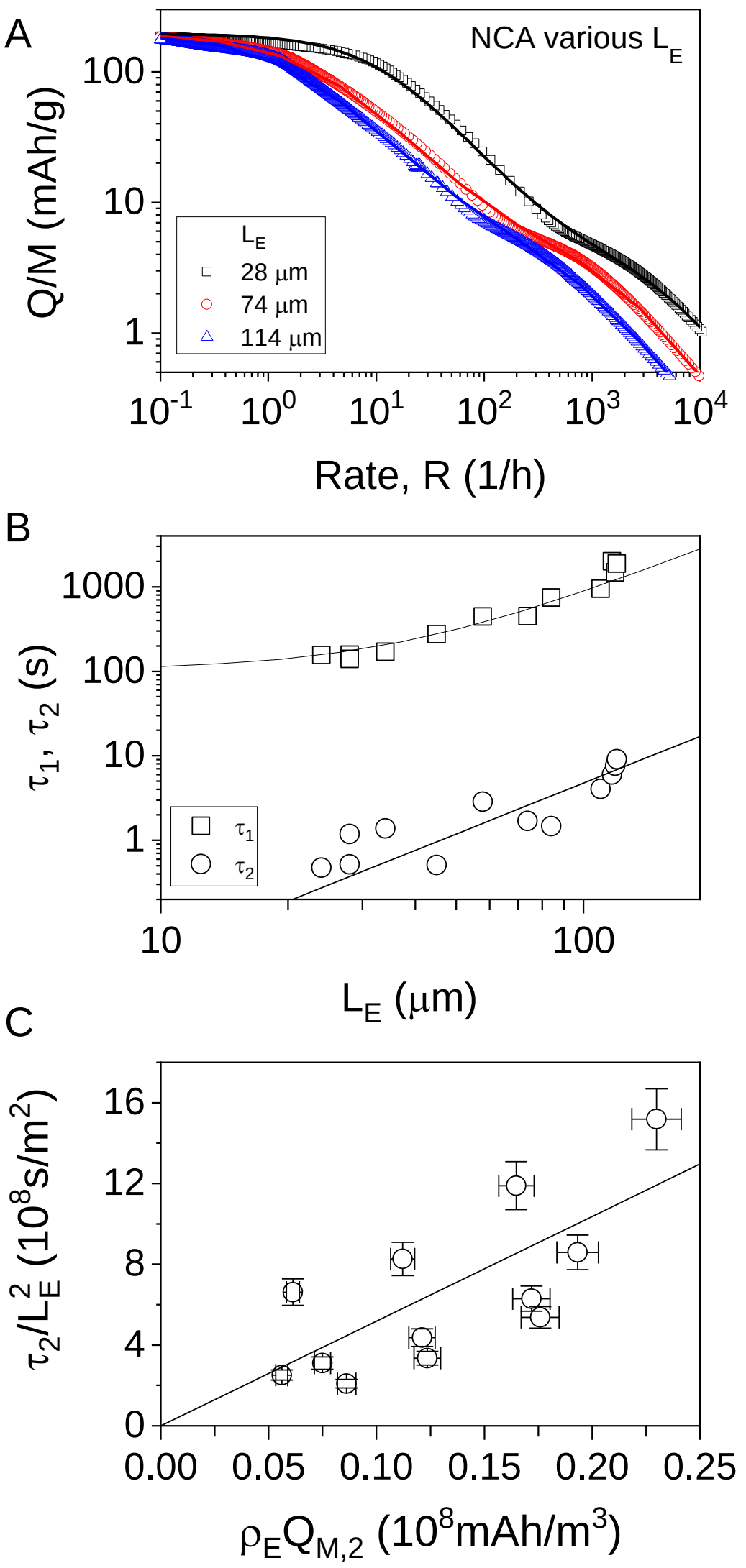

Figure 7: A) Capacity versus rate curves for NCA electrodes with three different thicknesses ($L_E$). The solid lines represent fits to equation 20. B) Time constants extracted from fitting data such as that in A, plotted versus electrode thickness, $L_E$. The lines are fits to equations 21a ($\tau_1$) and 21c ($\tau_2$). C) Fit data for the electrode thickness-dependent NCA data plotted as $\tau_2 / L_E^2$ vs. $\rho_E Q_{M,2}$. The line is a plot of equation 21c using the measured out-of-plane conductivity: $\sigma_E$=0.27 S/m.

| Q/M v R | Charge-discharge, low rate | | | CA, low rate | | |
|---|---|---|---|---|---|---|
| Material | $Q_M$ (mAh/g) | τ (h) | n | $Q_M$ (mAh/g) | τ (h) | n |
| NMC | 131(2) | 0.086(4) | 1.16(3) | 131.5(1) | 0.088(1) | 0.923(3) |
| Graphite | 396(20) | 0.32(5) | 0.81(3) | 379.5(1) | 0.286(1) | 0.937(1) |
| NCA | 196(9) | 0.20(3) | 0.92(3) | 194.5(2) | 0.243(1) | 0.874(3) |
| µSiGr | 2100(26) | 0.30(1) | 0.87(1) | 2052(2) | 0.315(1) | 0.932(1) |
| LTO | 134(2) | 0.24(1) | 1.04(3) | 132.7(1) | 0.218(1) | 0.926(4) |

Table 1: Fit parameters found by fitting the C/M v R data in figure 4 using equation (13). The numbers in brackets represent the fitting uncertainties associated with the last digit of the fit parameter.

| Q/M v $R_C$ | Charge-discharge, low rate | | | CA, low rate | | |
|---|---|---|---|---|---|---|
| Material | $Q_M$ (mAh/g) | τ (h) | n | $Q_M$ (mAh/g) | τ (h) | n |
| NMC | 125(1) | 0.204(7) | 1.80(13) | 131.0(2) | 0.075(1) | 0.872(8) |
| Graphite | 535(130) | 0.054(4) | 0.350(13) | 383.2(6) | 0.274(1) | 0.927(5) |
| NCA | 199(2) | 0.136(8) | 0.780(18) | 180.8(6) | 0.247(2) | 1.04(2) |
| µSiGr | 2380(100) | 0.122(3) | 0.55 (5) | 2128(6) | 0.221(1) | 0.776(6) |
| LTO | 130(2) | 0.402(9) | 1.25(6) | 131.1(2) | 0.176(2) | 0.839(8) |

Table 2: Fit parameters found by fitting the C/M v C-rate data in figure 3 using equation (15). The numbers in brackets represent the fitting uncertainties associated with the last digit of the fit parameter.

| Q/M v R | CA, low rate | | | CA, high rate | | |
|---|---|---|---|---|---|---|
| Material | $Q_{M,1}$ (mAh/g) | $\tau_1$ (h) | $n_1$ | $Q_{M,2}$ (mAh/g) | $\tau_2$ (h) | $n_2$ |
| NMC | 124.6(2) | 0.105(1) | 1.023(4) | 6.0(1) | 0.00097(5) | 0.970(3) |
| Graphite | 369.2(1) | 0.318(1) | 0.993(1) | 9.62(4) | 0.00083(5) | 0.930(1) |
| NCA | 189.3(2) | 0.265(2) | 0.935(4) | 3.9(1) | 0.00085(5) | 1.04(1) |
| uSiGr | 2036(2) | 0.316(1) | 0.949(1) | 4.75(8) | 0.00015(1) | 0.945(3) |
| LTO | 125.6(2) | 0.258(2) | 0.997(5) | 6.4(2) | 0.0022(1) | 0.939(4) |

Table 3: Fit parameters found by fitting the Q/M v R data in figure 6 using equation 20. The numbers in brackets represent the fitting uncertainties associated with the last digit of the fit parameter.

<table>
<tr><th rowspan="3">Materials</th><th rowspan="3">Electrode<br>(~4 mAh/cm$^2$)</th><th colspan="4">Test procedure</th></tr>
<tr><th>Step 1</th><th colspan="2">Step 2</th><th>Step 3</th></tr>
<tr><th>Formation step (cc mode), charge and discharge at same rate.</th><th>CC Charge</th><th>CA mode, discharge at lower cut-off potential.</th><th>GCD test</th></tr>
<tr><td>NCA</td><td>CNT: 1 wt%<br>PVDF: 5 wt%<br>NCA: 94 wt%</td><td rowspan="2">3.0-4.3 V, 1/20 C for 1 cycle, 1/10 C for 5 cycles.</td><td rowspan="2">Charge to 4.3 V at 1/20 C.</td><td rowspan="2">Set potential to 3.0 V.</td><td rowspan="2">Charge at 1/20 C to 4.3 V, and discharge at various rates to 3.0 V, 2 cycles for each rate.</td></tr>
<tr><td>NCM</td><td>Conductive Additive: 8%<br>Binder: 6%<br>NCM :86%</td></tr>
<tr><td>µSiGr</td><td>CNT: 7%<br>Graphite: 33%<br>Si: 60%</td><td>0.005-1.2 V,<br>1/20 C for 1 cycle,<br>1/10 C for 5 cycles.</td><td>Charge to 1.2 V at 1/10 C.</td><td>Set potential to 0.005 V.</td><td>Charge at 1/10 C to 1.2 V, and discharge at various rates to 0.005 V, 2 cycles for each rate.</td></tr>
</table>

| | | | | | |
|---|---|---|---|---|---|
| Graphite | Conductive Additive: 8%<br>Binder: 6%<br>Graphite :86% | 0.001−1.5 V,<br>1/20 C for 1 cycle,<br>1/10 C for 5 cycles. | Charge to 1.5 V at 1/10 C. | Set potential to 0.001 V. | Charge at 1/10 C to 1.5 V, and discharge at various rates to 0.001 V, 2 cycles for each rate. |
| LTO | Conductive Additive: 10%<br>Binder: 6%<br>LTO :84% | 1.0-2.5 V,<br>1/20 C for 1 cycle,<br>1/10 C for 5 cycles. | Charge to 2.5 V at 1/10 C. | Set potential to 1.0 V | Charge at 1/5 C to 2.5 V, and discharge at various rates to 1.0 V, 2 cycles for each rate. |

Table 4: Description of half-cell tests for different materials.

References

1. Armand, M.; Tarascon, J. M., Building better batteries. *Nature* **2008,** *451* (7179), 652-657.
2. Wong, L. L.; Chen, H. M.; Adams, S., Design of fast ion conducting cathode materials for grid-scale sodium-ion batteries. *Phys. Chem. Chem. Phys.* **2017,** *19* (11), 7506-7523.
3. Bonaccorso, F.; Colombo, L.; Yu, G. H.; Stoller, M.; Tozzini, V.; Ferrari, A. C.; Ruoff, R. S.; Pellegrini, V., Graphene, related two-dimensional crystals, and hybrid systems for energy conversion and storage. *Science* **2015,** *347* (6217).
4. Byeon, A.; Glushenkov, A. M.; Anasori, B.; Urbankowski, P.; Li, J. W.; Byles, B. W.; Blake, B.; Van Aken, K. L.; Kota, S.; Pomerantseva, E.; Lee, J. W.; Chen, Y.; Gogotsi, Y., Lithium-ion capacitors with 2D Nb2CTx (MXene) - carbon nanotube electrodes. *J. Power Sources* **2016,** *326*, 686-694.
5. Li, Q.; Wei, Q.; Zuo, W.; Huang, L.; Luo, W.; An, Q.; Pelenovich, V. O.; Mai, L.; Zhang, Q., Greigite Fe3S4 as a new anode material for high-performance sodium-ion batteries. *Chemical Science* **2017,** *8* (1), 160-164.
6. Park, S. H.; Tian, R. Y.; Coelho, J.; Nicolosi, V.; Coleman, J. N., Quantifying the Trade-Off between Absolute Capacity and Rate Performance in Battery Electrodes. *Adv. Energy Mater.* **2019,** *9* (33).
7. Tian, R. Y.; Park, S. N.; King, P. J.; Cunningham, G.; Coelho, J.; Nicolosi, V.; Coleman, J. N., Quantifying the factors limiting rate performance in battery electrodes. *Nature Commun.* **2019,** *10*.
8. McNulty, D.; Carroll, E.; O'Dwyer, C., Rutile TiO2 Inverse Opal Anodes for Li-Ion Batteries with Long Cycle Life, High-Rate Capability, and High Structural Stability. *Adv. Energy Mater.* **2017,** *7* (12).
9. Osiak, M.; Geaney, H.; Armstrong, E.; O'Dwyer, C., Structuring materials for lithium-ion batteries: advancements in nanomaterial structure, composition, and defined assembly on cell performance. *J. Mater. Chem. C.* **2014,** *2* (25), 9433-9460.
10. Yu, D. Y. W.; Donoue, K.; Inoue, T.; Fujimoto, M.; Fujitani, S., Effect of electrode parameters on LiFePO4 cathodes. *J. Electrochem. Soc.* **2006,** *153* (5), A835-A839.
11. Zhang, C. F.; Park, S. H.; Ronan, O.; Harvey, A.; Seral-Ascaso, A.; Lin, Z. F.; McEvoy, N.; Boland, C. S.; Berner, N. C.; Duesberg, G. S.; Rozier, P.; Coleman, J. N.; Nicolosi, V., Enabling Flexible Heterostructures for Li-Ion Battery Anodes Based on Nanotube and Liquid-Phase Exfoliated 2D Gallium Chalcogenide Nanosheet Colloidal Solutions. *Small* **2017,** *13* (34).

12. Zheng, H. H.; Li, J.; Song, X. Y.; Liu, G.; Battaglia, V. S., A comprehensive understanding of electrode thickness effects on the electrochemical performances of Li-ion battery cathodes. *Electrochim. Acta* **2012,** *71*, 258-265.
13. Heubner, C.; Nickol, A.; Seeba, J.; Reuber, S.; Junker, N.; Wolter, M.; Schneider, M.; Michaelis, A., Understanding thickness and porosity effects on the electrochemical performance of LiNi0.6Co0.2Mn0.2O2-based cathodes for high energy Li-ion batteries. *J. Power Sources* **2019,** *419*, 119-126.
14. Park, S. H.; King, P. J.; Tian, R. Y.; Boland, C. S.; Coelho, J.; Zhang, C. F.; McBean, P.; McEvoy, N.; Kremer, M. P.; Daly, D.; Coleman, J. N.; Nicolosi, V., High areal capacity battery electrodes enabled by segregated nanotube networks. *Nat. Energy* **2019,** *4* (7), 560-567.
15. Heubner, C.; Lammel, C.; Nickol, A.; Liebmann, T.; Schneider, M.; Michaelis, A., Comparison of chronoamperometric response and rate-performance of porous insertion electrodes: Towards an accelerated rate capability test. *J. Power Sources* **2018,** *397*, 11-15.
16. Higgins, T. M.; Coleman, J. N., Avoiding Resistance Limitations in High-Performance Transparent Supercapacitor Electrodes Based on Large-Area, High-Conductivity PEDOT:PSS Films. *ACS Appl. Mater. Interfaces* **2015,** *7* (30), 16495-16506.
17. Doyle, M.; Fuller, T. F.; Newman, J., MODELING OF GALVANOSTATIC CHARGE AND DISCHARGE OF THE LITHIUM POLYMER INSERTION CELL. *J. Electrochem. Soc.* **1993,** *140* (6), 1526-1533.
18. Fuller, T. F.; Doyle, M.; Newman, J., SIMULATION AND OPTIMIZATION OF THE DUAL LITHIUM ION INSERTION CELL. *J. Electrochem. Soc.* **1994,** *141* (1), 1-10.
19. Gallagher, K. G.; Trask, S. E.; Bauer, C.; Woehrle, T.; Lux, S. F.; Tschech, M.; Lamp, P.; Polzin, B. J.; Ha, S.; Long, B.; Wu, Q. L.; Lu, W. Q.; Dees, D. W.; Jansen, A. N., Optimizing Areal Capacities through Understanding the Limitations of Lithium-Ion Electrodes. *J. Electrochem. Soc.* **2016,** *163* (2), A138-A149.
20. Johns, P. A.; Roberts, M. R.; Wakizaka, Y.; Sanders, J. H.; Owen, J. R., How the electrolyte limits fast discharge in nanostructured batteries and supercapacitors. *Electrochem. Commun.* **2009,** *11* (11), 2089-2092.
21. Heubner, C.; Seeba, J.; Liebmann, T.; Nickol, A.; Borner, S.; Fritsch, M.; Nikolowski, K.; Wolter, M.; Schneider, M.; Michaelis, A., Semi-empirical master curve concept describing the rate capability of lithium insertion electrodes. *J. Power Sources* **2018,** *380*, 83-91.
22. Cornut, R.; Lepage, D.; Schougaard, S. B., Interpreting Lithium Batteries Discharge Curves for Easy Identification of the Origin of Performance Limitations. *Electrochim. Acta* **2015,** *162*, 271-274.
23. Zhang, L. J.; Peng, H.; Ning, Z. S.; Mu, Z. Q.; Sun, C. Y., Comparative Research on RC Equivalent Circuit Models for Lithium-Ion Batteries of Electric Vehicles. *Appl. Sci.-Basel* **2017,** *7* (10), 16.
24. Doyle, M.; Newman, J., Analysis of capacity-rate data for lithium batteries using simplified models of the discharge process. *J. Appl. Electrochem.* **1997,** *27* (7), 846-856.
25. Zhang, L. P.; Fu, J.; Zhang, C. H., Mechanical Composite of LiNi0.8Co0.15Al0.05O2/Carbon Nanotubes with Enhanced Electrochemical Performance for Lithium-Ion Batteries. *Nanoscale Res. Lett.* **2017,** *12*.
26. McNulty, D.; Buckley, D. N.; O'Dwyer, C., Comparative Electrochemical Charge Storage Properties of Bulk and Nanoscale Vanadium Oxide Electrodes. *J. Solid State Electrochem.* **2016,** *20* (5), 1445-1458.
27. Gewirth, A. A., Inorganic Electrochemistry: Theory, Practice and Application By Piero Zanello (University of Siena, Italy). Royal Society of Chemistry: Cambridge. 2003. xiv + 616 pp. $199.00. ISBN 0-85404-661-5. *J. Am. Chem. Soc.* **2004,** *126* (14), 4743-4744.
28. Ehrl, A.; Landesfeind, J.; Wall, W. A.; Gasteiger, H. A., Determination of Transport Parameters in Liquid Binary Lithium Ion Battery Electrolytes I. Diffusion Coefficient. *J. Electrochem. Soc.* **2017,** *164* (4), A826-A836.
29. Ong, M. T.; Verners, O.; Draeger, E. W.; van Duin, A. C. T.; Lordi, V.; Pask, J. E., Lithium Ion Solvation and Diffusion in Bulk Organic Electrolytes from First-Principles and Classical Reactive Molecular Dynamics. *J. Phys. Chem. B* **2015,** *119* (4), 1535-1545.

ToC fig

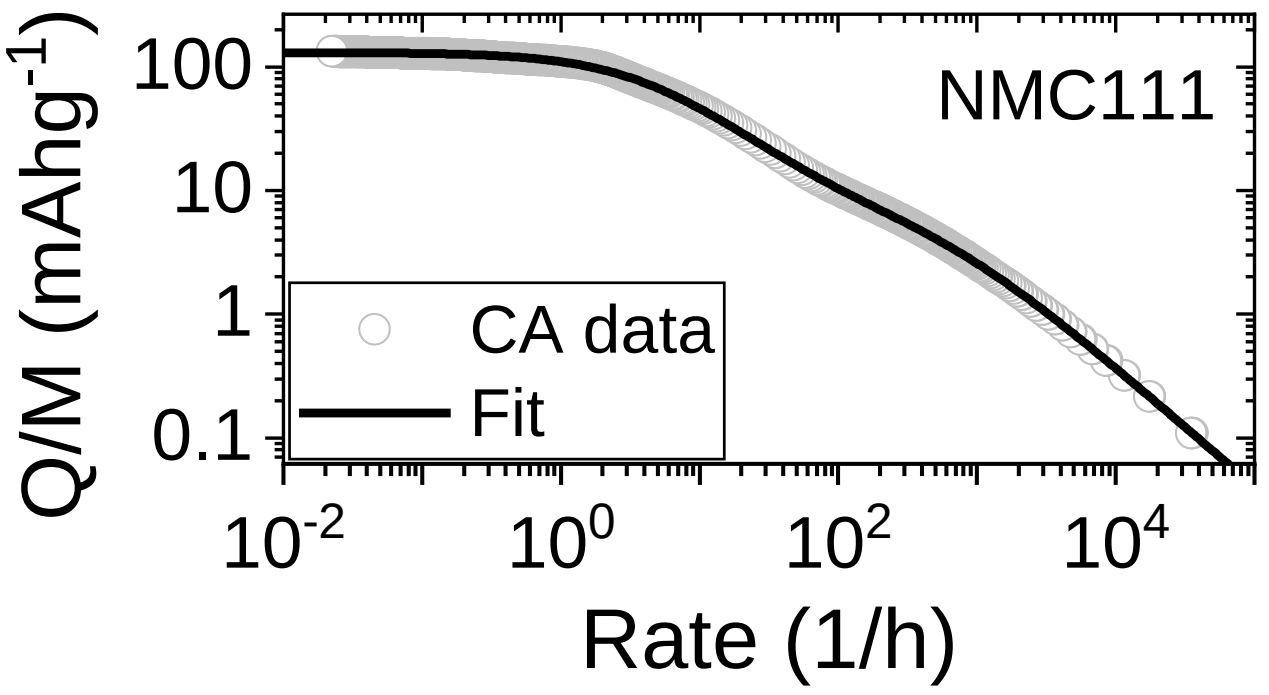